\documentstyle[graphicx]{report}

\renewcommand{\abstract}[1]{{ \footnotesize \noindent {\bf Abstract} #1 \\}}
\renewcommand{\author}[1]{\subsubsection*{\it#1}}
\newcommand{\address}[1]{\subsubsection*{\it#1}}

\begin{document}

\chapter*{Stars, gas and dust in elliptical galaxies\label{pipino1}}
\author{Antonio Pipino$^{a,b}$\footnote{email: axp@astro.ox.ac.uk},  Francesca Matteucci$^{b,d}$ and Thomas H. Puzia$^e$}
\address{$^a$Astrophysics, Oxford University, Denys Wilkinson Building, Keble Road, Oxford OX1 3RH, UK\\
$^b$Dipartimento di Astronomia, Universit\`a di Trieste, Via G.B. Tiepolo 11, 34100 Trieste, Italy\\
$^d$ INAF, Osservatorio Astronomico di Trieste, Via G.B. Tiepolo 11, 34100 Trieste, Italy\\
$^e$ Herzberg Institute of Astrophysics, 5071 West Saanich Road, Victoria, BC V9E 2E7, Canada, }


\abstract{I will present recent theoretical results on the formation and the high redshift assembly of spheroids. These findings have been obtained by utilising different and complementary techniques: chemodynamical models offer great insight in the radial abundance gradients in the stars; while state semi-analytic codes implementing a detailed treatment of the chemical evolution allow an exploration of the role of the galactic mass in shaping many observed relations. The results will be shown by following the path represented by the evolution of the mass-metallicity relation in stars, gas and dust. I will show how, under a few sensible assumptions, it is possible to reproduce a large number of observables ranging from the Xrays to the Infrared.
By comparing model predictions with observations, we derive a
picture of galaxy formation in which the higher is the mass of the
galaxy, the shorter are the infall and the star formation
timescales. Therefore, the stellar component of the most massive and luminous galaxies
might attain a metallicity $Z \ge Z_{\odot}$ in only 0.5 Gyr.
Each galaxy is created outside-in, i.e. the outermost regions accrete gas, form stars and
develop a galactic wind very quickly, compared to the central core in
which the star formation can last up to $\sim 1.3$ Gyr.  }

\section{Introduction}
Any model of galaxy evolution presented so far had to overcome the strong challenge
represented by the observational fact that elliptical galaxies show a remarkable
uniformity in their photometric and chemical properties,
one of the strongest constraints being the mass-metallicity relation (e.g. Carollo
et al. 1993, Davies et al. 1993).
The first proposed scenario of elliptical formation was the so-called
monolithic collapse scenario (e.g. Larson, 1974). In this framework,
ellipticals are assumed to have formed at high redshift as a result of a rapid collapse of a gas cloud.
This gas is then rapidly converted into stars by means of a very strong burst,
followed by a galactic wind powered by the energy injected into the interstellar medium (ISM) by
supernovae (SNe) and stellar winds.
The wind carries out the residual gas from the galaxies, thus inhibiting further star formation.
In this way the mass-metallicity (i.e. more massive a galaxies have the higher
metal content in stars and gas) relation
could be easily explained in terms of metallicity sequences, namely the more massive objects
develop the wind later (due to their deeper potential wells) and, thus, have more time
to enrich their stellar generations.
This scenario has been recently reviewed and modified by Pipino \& Matteucci (2004, PM04) in order to
take into account that ellipticals show an increasing Mg/Fe abundance ratio in the stars
as a function of galactic mass (Faber et al. 1992). Due to the different nucleosynthesis leading to the production
of Mg (by type II SNe, on short timescales) and Fe (by type Ia SNe, on longer timescale), the 
Mg/Fe-mass relation implies that the more massive objects should have formed faster than the less massive
ones (see Matteucci, this book, and references therein). Pipino \& Matteucci (2004) implemented an infall term in the chemical evolution
equation in order to simulate the creation of galaxies and found that 
most of the photo-chemical observables, including the Mg/Fe-mass relation can be 
reproduced in a scenario in which
the more massive galaxies formed faster and with a much more efficient star formation process
with respect to the low mass objects.

PM04 suggested that a single galaxy should form outside-in, namely the
outermost regions form earlier and faster with respect to the central
parts.  A natural consequence of this
model and of the time-delay between the production of Fe and that of Mg
is that the mean [Mg/Fe] abundance ratio in the stars should
increase with radius. 
Pipino et al. (2006, PMC06) compared PM04 best model results with the very recent
observations for the galaxy NGC 4697 (Mendez et al. 2005), and found
them in excellent agreement.

Metallicity gradients, in fact, are characteristic of the stellar populations
inside elliptical galaxies. Evidences come from the increase of
line-strength indices (e.g. Carollo et al., 1993; Davies et al., 1993;
Trager et al., 2000) and the reddening of the colours (e.g. Peletier
et al. 1990) towards the centre of the galaxies.  The study of such
gradients provide insights into the mechanism of galaxy
formation, particularly on the duration of the chemical enrichment
process at each radius.  Metallicity indices, in fact, contain
information on the chemical composition and the
age of the simple stellar populations (SSPs) inhabiting a given
galactic zone.

\section{The model}
The chemical code adopted here is described in full detail 
in PM04 and PMC06, where we address the reader for more details. 
This model is characterized by:
Salpeter (1955) IMF, Thielemann et al. (1996) yields for massive stars,
Nomoto et al. (1997) yields for type Ia SNe and 
van den Hoek \& Groenewegen (1997) yields for low-
and intermediate-mass stars (the case with $\eta_{AGB}$ varying with metallicity). 
Here we present our analysis of a $\sim 10^{11}M_{\odot}$ galaxy (PM04 model IIb), considered
representative of a typical elliptical, unless otherwise stated.

The model assumes that the galaxy assembles by merging of gaseous
lumps (infall) on a short timescale and suffers a strong star burst
which injects into the interstellar medium a large amount of
energy able to trigger a galactic wind, occurring at different
times at different radii.  After the development of the
wind, the star formation is assumed to stop and the galaxy evolves
passively with continuous mass loss.

\section{Results and discussion}

\subsection{The mass-metallicity relation evolution}

\begin{figure*}
\includegraphics[width=2.6in,height=2.4in]{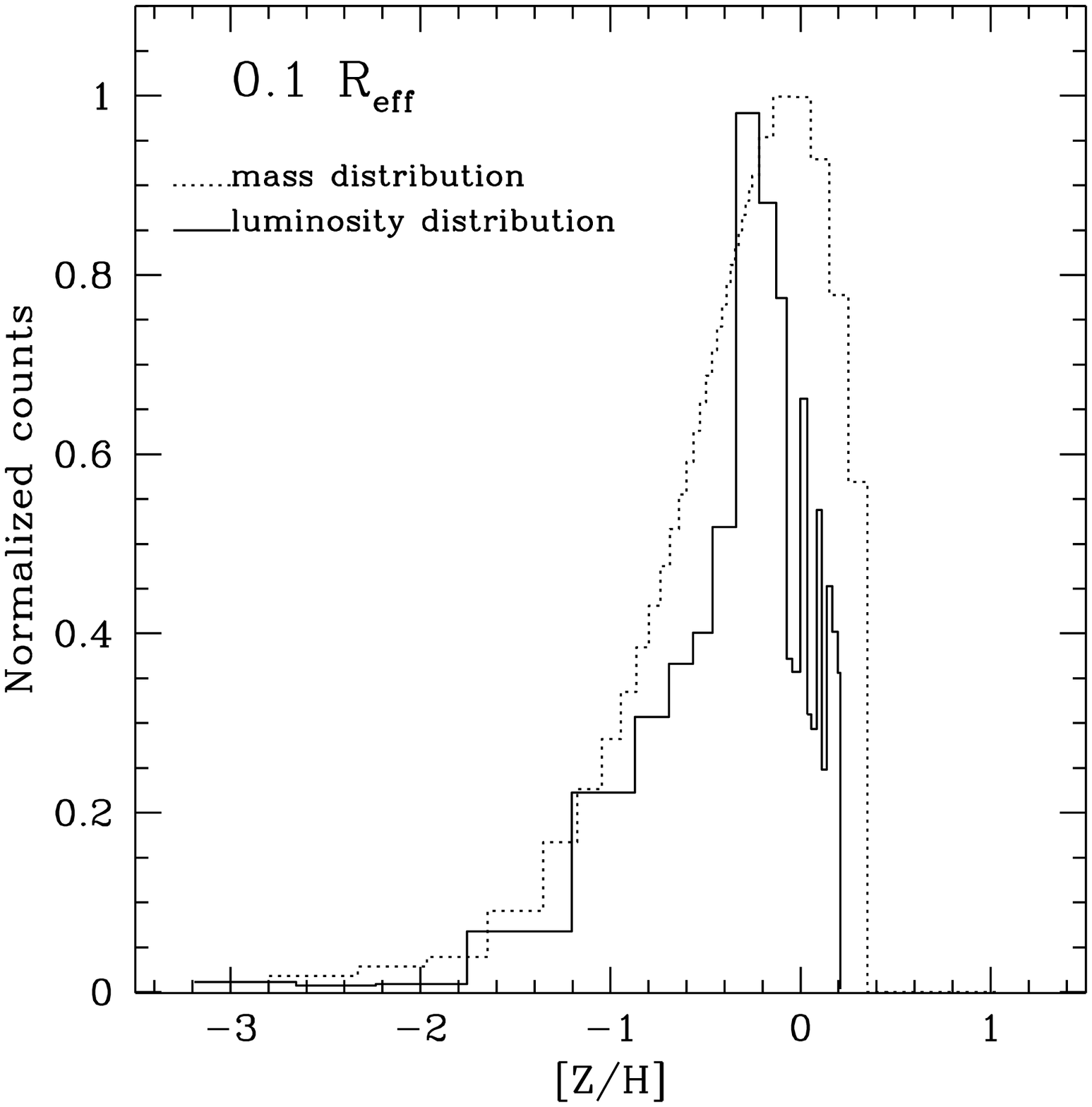}
\includegraphics[width=2.6in,height=2.4in]{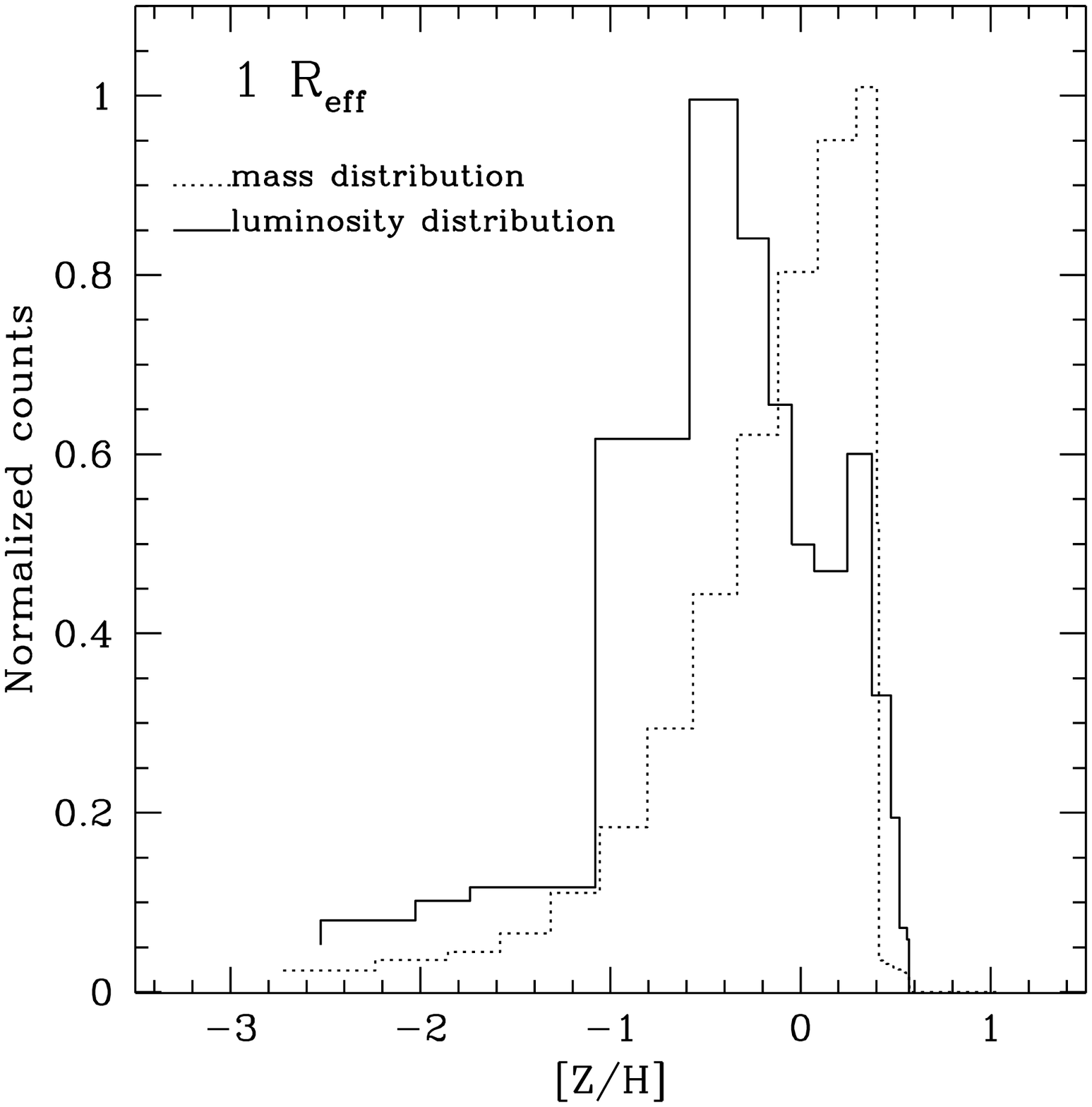}
\caption{Stellar metallicity distribution distributions for  
[Z/H] in luminosity (solid line) and mass (dotted line).
\emph{Upper panel}: values at $0.1 R_{eff}$.
\emph{Lower panel}: values at $1 R_{eff}$.
The plots are presented in the same scale in order to better
appreciate the differences among the different distributions.
}
 \label{Gdwarf}
\end{figure*}

From the comparison between our model predictions (Fig.~\ref{Gdwarf}) and the observed
stellar metallicity distribution diagrams derived at different radii 
by Harris \& Harris (2002, see their fig. 18) for the elliptical galaxy NGC 5128, we can derive some general considerations.
The qualitative agreement is remarkable: we can explain the slow rise in the [Z/H]-distribution
as the effect of the infall, whereas the sharp truncation at high metallicities is
the first direct evidence of a sudden and strong wind which stopped the star formation.
The suggested outside-in formation process reflects in a more asymmetric shape of the stellar metallicity distribution diagram
at larger radii, where the galactic wind occurs earlier (i.e. closer to the peak of the star formation rate),
with respect to the galactic centre.

\begin{figure*}
\centering
\includegraphics[width=3.5in,height=2.4in]{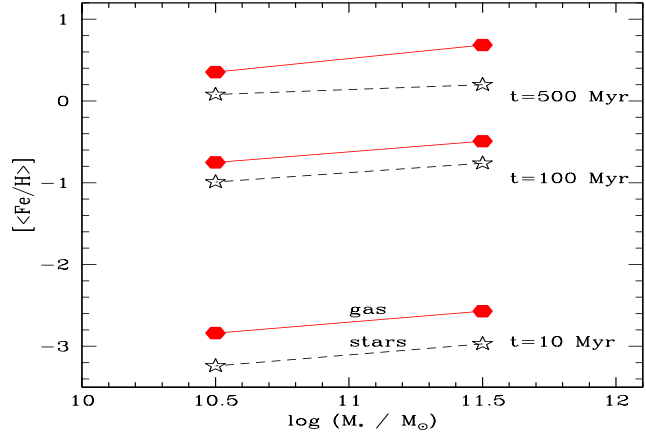}
\caption{The temporal evolution of mass-metallicity relation for the two studied galactic components (stars and gas).}
 \label{massmet}
\end{figure*}

From a quantitative point of view, properties suchs as the stellar metallicity distribution of the composite
stellar populations (CSPs) inhabiting 
the galactic core, allow us to study
the build-up of mass-metallicity relation, which
is tipically inferred from the spectra taken at $\sim 0.1$ effective radius.
In Fig.~\ref{massmet} we plot the time evolution of the mass-metalliticity relation
in stars (which reflect the average chemical enrichment of the galactic core as seen at the present day; dashed line) and in the gas
(which, instead, is closer to the composition of the youngest SSP, thus being more indicative of a high redshift object; solid line).
The mean Fe abundance in the stellar component can reach the solar value in only 0.5 Gyr,
making ellipticals among the most metal-rich objects of the universe. 
FOr the star formation histories of these objects see Matteucci (this book).
For the metals locked-up in dust we refer the reader to Calura, Pipino \& Matteucci (2007) and Calura (\ref{calura}, this book).

It is intriguing that the Lyman-Break Galaxies observed by Verma et al. (2007) at a redshift of 5
have assembled probably less than 1/10 of their final mass in stars (at a rate of $\sim 40 \rm M_{\odot}/ yr$), they have an
average age which is lower than 100 Myr and an inferred gas metallicity below Z = 0.01 Z$_{\odot}$.
Their finding is in remarkable agreement
with our expectations for a present-day elliptical caught in the act of the build-up of the low-metallicity tail of the stellar metallicity distribution curve
at very high redshift.

On the other hand, at variance with the stellar metallicity distribution diagrams as a function of [Z/H] (and [Fe/H]),
abundance ratios such as [$\alpha$/Fe] have  narrow and almost symmetric distributions. This means that, also
from a mathematical point of view, the [$<\alpha/Fe>$] ratio
are representative of the whole CSP (PMC06). The robustness of the [$\alpha$/Fe] ratios as constraints for
the galactic formation history is testified by the fact that 
$[\rm <\alpha/Fe>]\simeq [\rm <\alpha/Fe>_V]$, having very similar
distributions.
In particular, we find that the skewness parameter is much larger for the [Z/H] and [Fe/H]
distributions than for the case of the [$\alpha$/Fe] one, by
more than one order of magnitude. Moreover, the asymmetry
increases going to large radii (see Fig.~\ref{Gdwarf}, lower panel), up to a factor
of $\sim$7 with respect to the inner regions. Therefore, it is not surprising that
the $[\rm <Z/H>]$ value does not
represent the galaxy at large radii,
and hence, we stress that care should be taken when
one wants to infer the real abundances of the stellar components
for a galaxy by comparing the observed indices (related
to a CSP) with the theoretical ones (predicted
for a SSP). Only the comparison based on the $[\rm <\alpha/Fe>]$ ratios
seems to be robust.


Another possible source of discrepancies is the fact
that luminosity-weighted averages (which are more closely related
to the observed indices) and mass-weighted averages (which
represent the real distributions of the chemical
elements in the stellar populations) might differ more in the most external
zones of the galaxy (compare the panels in Fig.~\ref{Gdwarf}).
All these considerations result in the fact that the chemical
abundance pattern used by modellers to build their SSPs, might
not necessary reflect the real trends.
Therefore, the interpretation of line strenght indices
in terms of abundances, can be seriously flawed (see PMC06 for further details).
We refer to Pipino et al. (2007b, see also \ref{pipino2} ,this book) where, the build-up
of the metallicity gradients is thorougly studied.

\subsection{Globular cluster systems in ellipticals: a different point of view}

The analysis of the radial variation in the CSPs inhabiting elliptical galaxies seems to be promising as
a powerful tool to study ellipticals. Pipino, Puzia \& Matteucci (2007),
make use of the stellar metallicity distributions predicted by PMC06 to explain the multimodality in the globular
cluster (GC) metallicity distribution as well as their high $\alpha$ enhancement (Puzia et al. 2006, P06). In particular, they 
show that the GC distribution as function of [Fe/H] for the whole galaxy can be constructed
simply by combining distributions as those of Fig.~\ref{Gdwarf} (typical of different radii), once they had been rescale by means of a suitable
function (of time and metallicity) which links the global star formation rate
to the globular cluster creation. 

\begin{figure*}


\includegraphics[width=3.5in,height=2.4in]{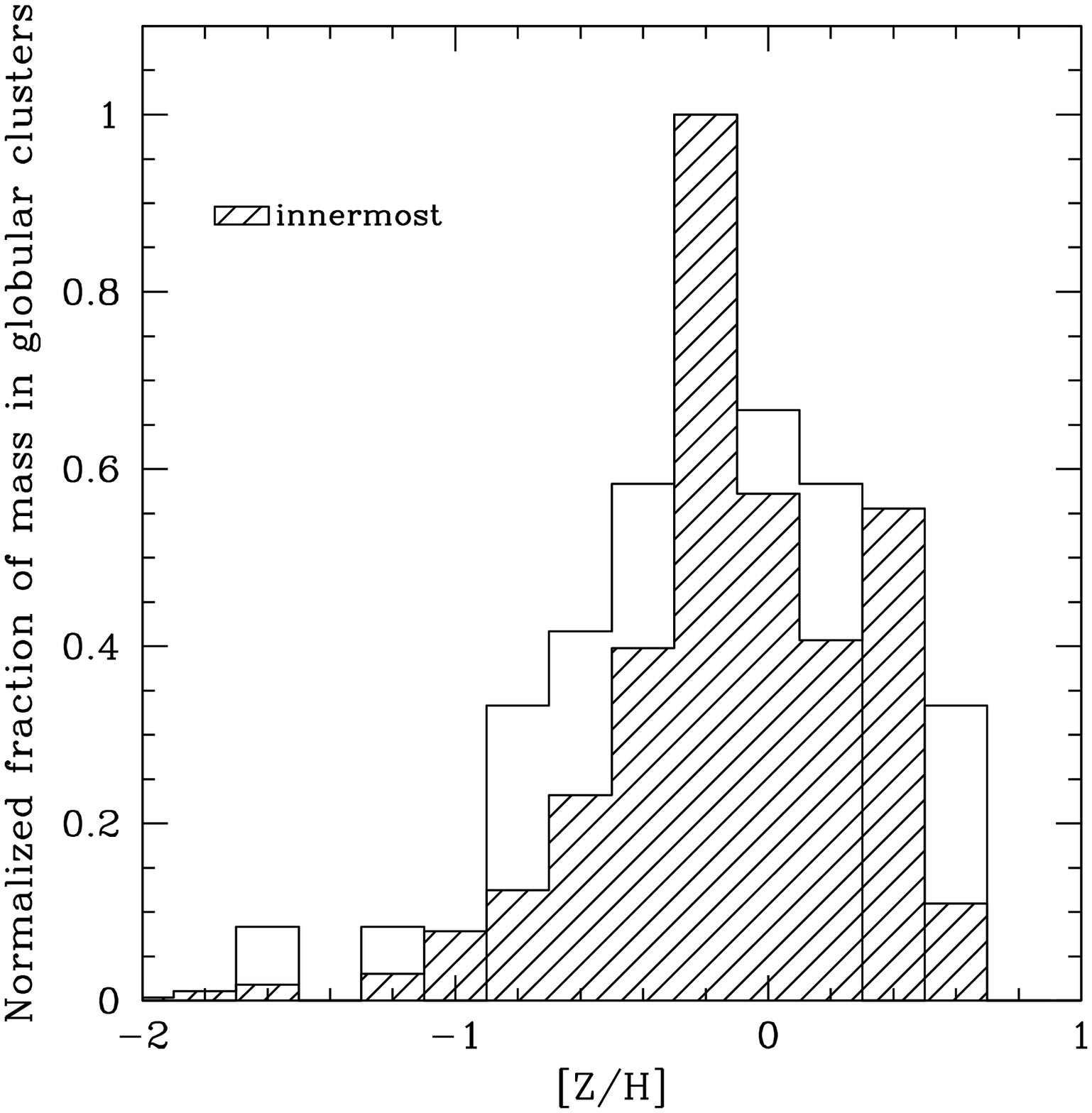}
\includegraphics[width=3.5in,height=2.4in]{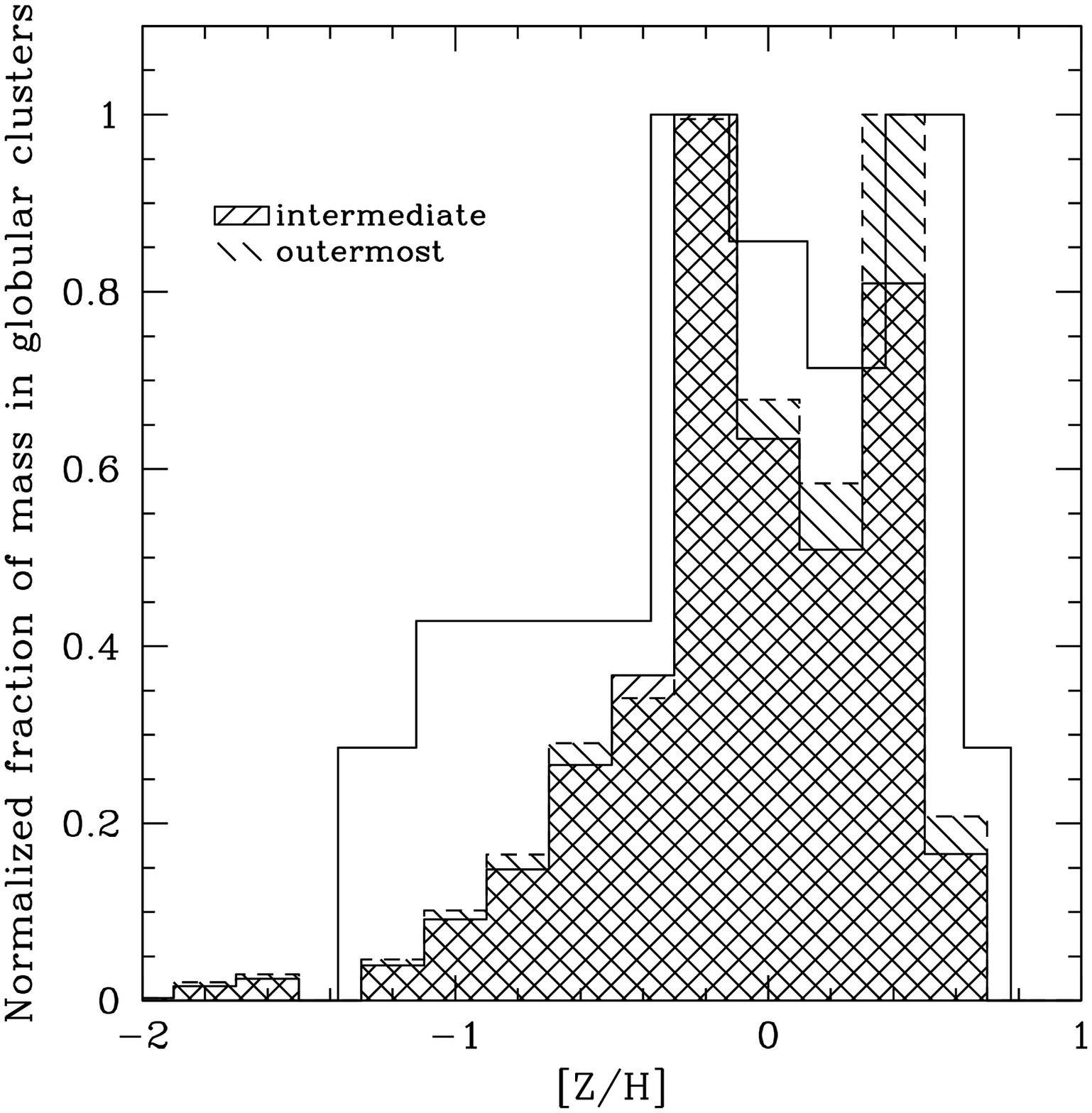}

\caption{Predicted globular-cluster metallicity distribution by mass as a function of [Z/H] for three different radial
compositions. The upper panel shows both model
predictions and observations related to the central part of an elliptical
galaxy. The lower panel shows the same quantities for cluster populations
residing at $r \ge R_{\rm eff}$. Solid empty histograms:
observational data taken as sub-samples of the P06 compilation, according
to the galactic regions presented in each panel.}

\label{GD_mass}

\end{figure*}

In order to plot the different cases on the same scale we normalize each
GC-metalliticy distribution function by its maximum value. In the upper panel of
Figure~\ref{GD_mass} the shaded histogram represents the innermost
population. Our predictions match the data very well, especially in the
metal-rich slope and the mean of the distribution. The same happens for
the \emph{pure core} populations, which shows how the GCS might be used to
probe the CSP in ellipticals. It should be remarked that a second peak
centered at super-solar metallicity appears in the distribution predicted
by our models, although not evident in the data of the particular radial
sub-sample. The lower panel of Figure~\ref{GD_mass} illustrates model
predictions which are more representative of the galaxy as a whole (either
at $1 R_{\rm eff}$, i.e. the \emph{intermediate} population, or at several
effective radii, the \emph{outermost} population), and we consider them as
the fiducial case. These two cases look quite similar to each other and
have clear signs of bimodality in remarkable agreement with the
spectroscopic data (solid empty histogram, sub-sample of the P06 data with
$r \ge R_{\rm eff}$). { A {Kolmogorov-Smirnov} test returns $>\!99\%$
probability that both model predictions and observations are drawn from
the same parent distribution in the upper panel of Figure~\ref{GD_mass}.
The lower panel statistics gives a lower likelihood of $98.4\%$ that both
distributions have the same origin, which is mainly due to the observed
excess of metal-poor GCs at large galactocentric radii compared to the
model predictions.} The prediction of a super-solar metallicity globular
cluster sub-population is entirely new and a result of the radially
varying and violent formation of the parent galaxy. Moving to the
low-metallicity tail, we predict slightly fewer low metallicity objects
than expected from observations , which is attributed to the lack of GC accretion from
metal-poor satellite galaxies in our model.

\begin{figure*}


\includegraphics[width=3.5in,height=2.4in]{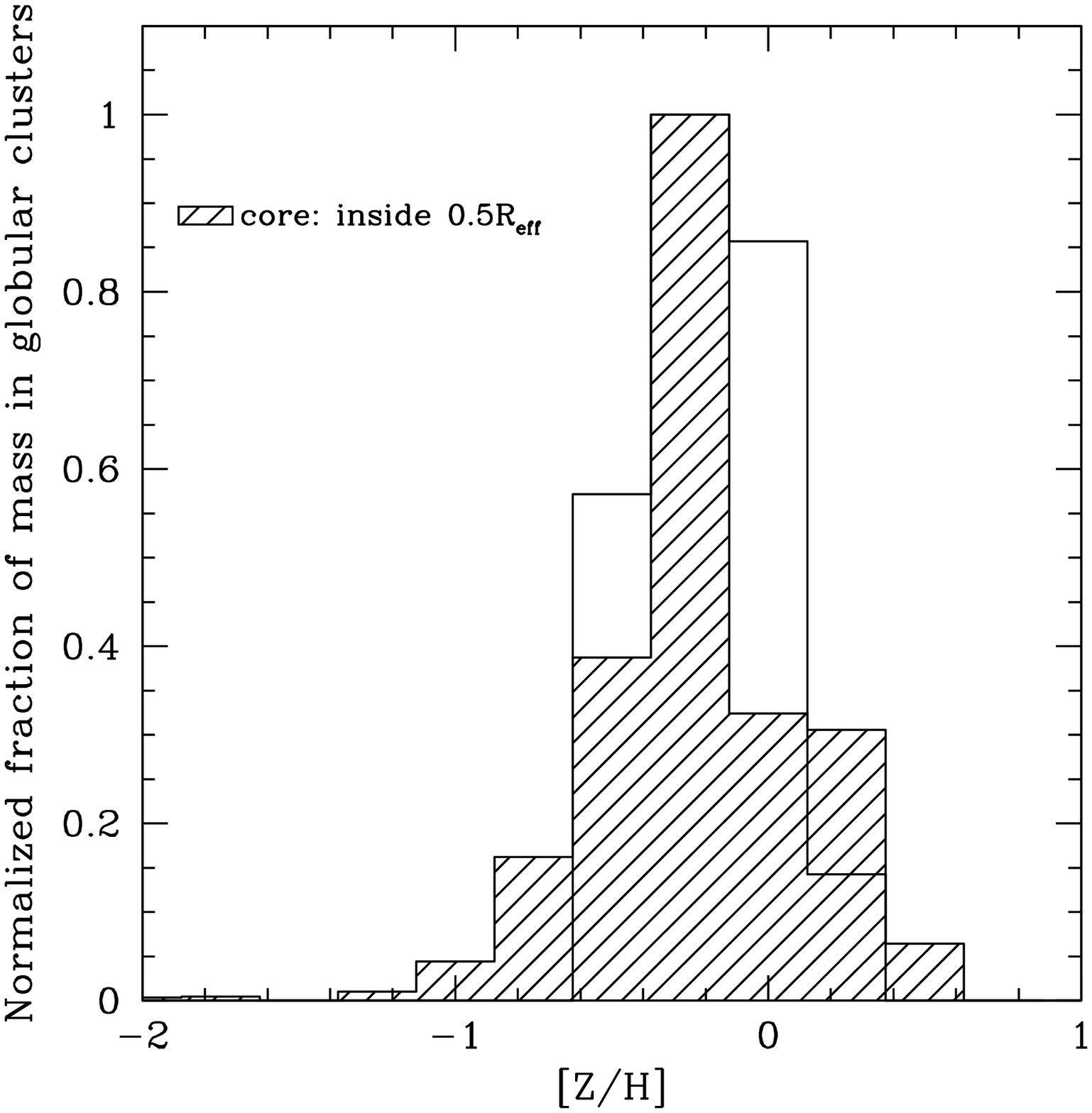}
\includegraphics[width=3.5in,height=2.4in]{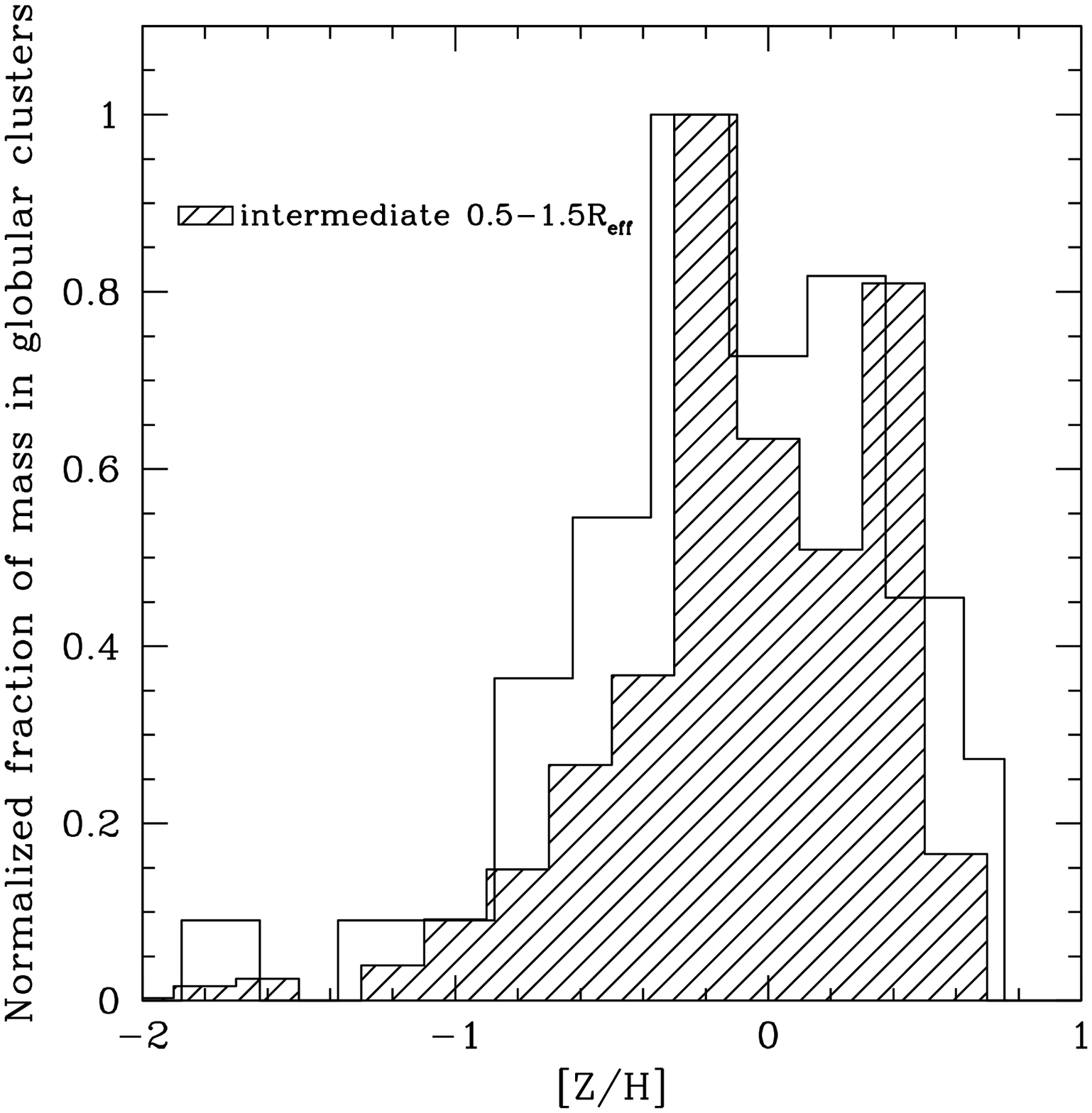}

\caption{Predicted globular-cluster metallicity distribution  by mass as a function of [Z/H] for two different projected
galactocentric radii. The upper panel shows both model predictions and
observations related to the \emph{pure core} of an elliptical galaxy
. The lower panel shows the same quantities
for cluster populations residing either at $0.5R_{\rm eff} < r <1.5 R_{\rm
eff}$. Solid empty histograms:
observational data taken as sub-samples of the P06 compilation, according
to the galactic regions presented in each panel.}
\label{GD_mass2}

\end{figure*}

In Figure~\ref{GD_mass2} (upper panel) we show the results for a \emph{pure
core} GCs, namely one in which we adopt $f_{red}\!:\!f_{blue}=1\!:\!0$. In
this quite extreme case the observed GCs have been selected with radius $r
< 0.5 R_{\rm eff}$. The histogram reflects the shape of a stellar metallicity distribution
diagram expected for a typical CSP inhabiting the galactic core. This
finding is particularly important, because it might offer the opportunity
to resolve the SSPs in ellipticals, at variance with data coming from the
integrated spectra which deal with luminosity-weighted quantities. Whereas
in Figure~\ref{GD_mass2} (lower panel), the intermediate population is
compared to a sub-sample of P06 GCs with $0.5 < r < 1.5 R_{\rm eff}$. This
is to show that the multimodality is not an artifact due to the particular
radial binning adopted in this paper.

Since the GC populations trace the properties of galactic CSPs in our scenario,
we predict an increase of the mean metallicity of the cluster systems with
the host galaxy mass, which closely follows the mass-metallicity relation
for ellipticals.This has been observed in the GCSs of Virgo cluster early-type galaxies (Peng et al. 2006, ApJ
639, 95). Moreover, we expect that a major fraction of the GCs
(i.e.~those born inside the galaxy) follows an age-metalliticity
relationship, in the sense that the older ones are also more
$\alpha$-enhanced and more metal-poor.

Neither a need of an enhanced GC formation during mergers
nor a strong role of the accretion of exteranl objects, seems to be required
in order to explain the different features of the GC metallicity distributions.
Since globular clusters are the closest approximation of a SSP, we expect that this technique
will be very helpful to probe the properties of the stellar populations in spheroids,
thus avoiding the uncertainties typical of the analysis based on their integrated spectra.

\section{Concluding remarks}

A detailed study of the chemical properties of the CSPs inhabiting elliptical
galaxies as well as the change of their properties as a function of both time and radius, allow us to 
gather a wealth of information. Our main conclusions are:

\begin{itemize}
\item Both observed and predicted stellar metallicity distribution
for ellipticals show a sharp truncation at high metallicities that, in the light of our models, might
be interpreted as the first direct evidence for the occurrence of the galactic
wind in spheroids.
\item The stellar component of the most massive and luminous galaxies
might attain a metallicity $Z \ge Z_{\odot}$ in only 0.5 Gyr.
\item PM04's best model prediction of increasing [$\rm <\alpha/Fe>$] ratio
with radius is in very good agreement with the observed
gradient in [$\alpha$/Fe] of NGC 4697.
This strongly suggests an outside-in galaxy formation scenario
for elliptical galaxies that show strong gradients (see also Pipino
et al. 2007b and \ref{pipino2}, this book).
\item By comparing the radial trend of [$\rm <Z/H>$] with the 
\emph{observed} one,
we notice a discrepancy which is due to the fact that a
CSP behaves in a different way with respect to a SSP.
In particular the predicted gradient of [$\rm <Z/H>$] is flatter
than the observed one at large radii.
Therefore, this should be taken into account when 
estimates for the metallicity of a galaxy are derived
from the simple comparison between the observed line-strength index
and the predictions for a SSP, a method currently adopted in the literature.
\item Abundance ratios such as [Mg/Fe] are less affected by the discrepancy
between the SSPs and a CSP, since
their distribution functions are narrower and more symmetric.
Therefore, we stress the importance of such a ratio as the
most robust tool to estimate the duration of the galaxy
formation process.
\item We show that the observed multi-modality in the globular cluster
metallicity distributions can be, at least partly, ascribed to the radial variation in the
underlying stellar populations in giant elliptical galaxies. In
particular, the observed globular cluster systems are consistent with a
linear combination of the globular cluster sub-populations inhabiting
different galactocentric radii projected on the sky. 

\item A new prediction of our models, which is in astonishing agreement
with the spectroscopic observations, is the presence of a super-solar
metallicity mode that seems to emerge in the most massive elliptical
galaxies. In smaller objects, instead, this mode disappears quickly with
decreasing stellar mass of the host galaxy.

\item Since in our scenario the GCs properties sample the galactic CSPs,
we predict an increase of the mean metallicity and mean [$\alpha/Fe$] of cluster systems with
the host galaxy mass, which closely follows the scaling relations for ellipticals. Moreover, we expect that a major fraction of the GCs
(i.e. those born inside the galaxy) follows an age-metalliticity
relationship, in the sense that the older ones are also more
$\alpha$-enhanced and more metal-poor.
\end{itemize}

A.P. thanks the Organizers for having provided financial support for attending the conference.

\end{document}